\documentclass{emulateapj}  
\usepackage{epsf}
\usepackage{latexsym}  
\usepackage{dcolumn}
\input{epsf}
\usepackage{epsfig}
\usepackage{natbib}

\newcommand{\beq}{\begin{equation}}
\newcommand{\eeq}{\end{equation}}
\def\msun{{\rm M}_\odot}

\begin{document}

\bibstyle{aas}

\def\msun{{\rm M}_\odot}

\shorttitle{Sharpening SZ Precision}
\shortauthors{Shaw et al.}

\author{ 
Laurie D. Shaw,\altaffilmark{1} Oliver Zahn,\altaffilmark{2}
Gilbert P. Holder,\altaffilmark{1} Olivier Dor\'e\altaffilmark{3}
}

\altaffiltext{1}{Department of Physics, McGill University, Montreal QC, Canada, H3A 2T8}
\altaffiltext{2}{Berkeley Center for Cosmological Physics, Department
of Physics, University of California, and Lawrence Berkeley National
Labs, Berkeley, CA 94720}
\altaffiltext{3}{Canadian Institute for Theoretical Astrophysics, 60
St. George, University of Toronto, Toronto, ON, Canada M5S3H8}

\email{lds@physics.mcgill.ca}

\title{Sharpening the Precision of the Sunyaev-Zel'dovich Power Spectrum}

\begin{abstract}
Using both halo model calculations and a large sample of simulated SZ
maps, we demonstrate that high-mass clusters add significant
non-Gaussian variance to measurements of the SZ power spectrum
amplitude. The difficulty in correctly accounting theoretically for
the contribution of these objects to the uncertainty in $C_\ell$ leads
to a reduced sensitivity to $\sigma_8$. We show that a simple solution
is to mask out the brightest clusters in the map before measuring the
power spectrum. We demonstrate that fairly conservative masking can
reduce the variance and Gaussianize the statistics significantly, thus
increasing the sensitivity to cosmological parameters. Choosing which
objects to mask is non-trivial; we found that using a fixed sky
density produced a well-defined and well-behaved estimate that can
easily be applied to real maps. For example, masking the 10 (90)
brightest clusters in a 100 deg$^2$ SZ map will improve the
sensitivity to $C_\ell$ by a factor of two at $\ell = 1000$ (2000) and
1.5 at $\ell = 2000$ (4000). We show that even in the presence of
astrophysical foregrounds (primary CMB and point sources) and
instrument noise, one can increase the precision on measurements of
$\sigma_8$ by masking up to 0.9 clusters deg$^{-2}$.
\end{abstract}

\section{INTRODUCTION}

The Sunyaev-Zel'dovich (SZ) effect has long been recognized as a
powerful tool for probing the physics of the intra-cluster medium,
large-scale structure formation and the dark energy equation of state
\citep{Birkinshaw:99, Carlstrom:02}. Simple but robust analytical
arguments \citep{Barbosa:96, Holder:01} and hydrodynamical simulations
of clusters \citep{White:02, Motl:05, Nagai:06} have indicated that
the integrated SZ flux (the integral of the temperature decrement
across the surface area of a cluster) should correlate tightly with
cluster mass. Combined with a greater sensitivity than optical or
X-ray surveys to high redshift objects, this makes SZ-selected cluster
samples well suited for measuring the evolution of the cluster mass
function over a wide range in redshift. Experiments such as the
Atacama Cosmology Telescope \citep{Kosowsky:03} and the South Pole
Telescope \citep{Ruhl:04} are currently surveying the microwave sky to
develop large catalogs of galaxy clusters that are uniformly selected
by the SZ flux.

However, in order to use cluster samples for this purpose, it is
necessary to have a good understanding of both the selection function
of the survey \citep{Melin:06, Schafer:07} and of the mapping between
the measured integrated SZ flux and cluster mass (and how this evolves
with redshift). Besides the slope and normalization of the flux-mass
(Y-M) relation, one must also gain some measure of the scatter around
the mean relation in order to be able to differentiate between
different dark energy cosmologies \citep{Battye:03, Lima:05}. A
detailed characterization of the Y-M relation may require X-ray
follow-up of a representative sub-sample of SZ-selected clusters
\citep{Majumdar:03}. Complete optical coverage of surveyed fields is
necessary for obtaining cluster redshifts and to help determine the
selection function. Thus, although there is much power in this method,
much work is required before it can be applied to its full potential.

The SZ effect can also be detected as a secondary anisotropy in the
CMB temperature anisotropy power spectrum, appearing as `excess power'
(over the predicted primary anisotropy signal) on angular scales below
several arc-minutes. In principle, measuring the SZ angular power
spectrum is an easier task than serendipitous cluster detection as one
is searching for temperature fluctuations (on a given angular scale)
to which clusters over a wide mass and redshift range contribute
significantly, without the need to resolve individual clusters at high
signal-to-noise. 

Predictions for the SZ power spectrum amplitude $C_{\rm \ell,sz}$
(henceforth $C_\ell$) can be made using the halo model
\citep[e.g. ][]{Cooray:01} and estimates of the radial pressure
profile of intra-cluster gas. Assuming that the cluster gas resides in
hydrostatic equilibrium in the potential well of the host dark matter
halo (with an additional steepening of the gas density profile in the
outer regions), \citet{Komatsu:02} demonstrated that the ensemble
averaged power spectrum amplitude $\bar C_\ell$ has an extremely
sensitive dependence on $\sigma_8$, where $\bar C_\ell \propto
\sigma_8^7(\Omega_b h)^2$. Although there remains a certain ambiguity
in the exact amplitude and shape of the predicted signal due to
uncertainties in cluster gas physics (and understanding of the
contribution of sub-mm and synchrotron sources to the power spectrum
provides an added complication), the SZ angular power spectrum
represents a robust observable with which competitive constraints on
$\sigma_8$ can be obtained.

Rather than attempting to resolve individual clusters, small-scale CMB
experiments have focused on detecting SZ power at arc-minute scales
and measuring the amplitude of $C_\ell$ and thus $\sigma_8$. BIMA
surveyed 0.2 square degrees on the sky measuring a fluctuation
amplitude of $\Delta T^2=C_\ell \ell (\ell+1)/2 \pi =
220^{+140}_{-120} \mu K^2$, from which they inferred a value of
$\sigma_8=1.03^{+0.20}_{-0.29}$ \citep{Dawson:2006qd} and CBI recently
measured an excess of $1.6 \sigma$ above $\sigma_8=0.8$
\citep{Sievers:09}. Clearly there is some tension between the value of
$\sigma_8$ inferred from these experiments and the current value of
$\sigma_8 \sim 0.8$ favored by WMAP and X-ray cluster number counts
\citep{Dunkley:09, Vikhlinin:09}. However, we note that recent results
from SZA \citep{Sharp:2009vq} and QUaD \citep{Friedman:2009dt}
measurements of the SZ power spectrum infer a value of $\sigma_8$ that
is more consistent with those derived from other methods.

One possible explanation for this discrepancy is that the ICM gas
models used to calculate the SZ power spectrum do not include or
correctly capture the relevant physics that determine the shape and
amplitude of pressure profiles, resulting in a large modeling
uncertainty in the amplitude of $C_\ell$. Analytic arguments
\citep{Holder:00} and hydrodynamical simulations \citep{daSilva:01,
White:02} suggest that the inclusion of more detailed physics
(radiative cooling, star formation and energy feedback) only
moderately influence the amplitude of the power spectrum. This is
partly because these effects mostly influence the temperature and
density of gas in cluster cores, whereas it is the gas at larger radii
that contributes most to $C_\ell$ \citep{Komatsu:02}. However, to date
there has not yet been a systematic comparison between simulated and
analytic power spectra, partly due to the limited total area of SZ
maps that can be produced from the hydrodynamical simulations, and so
the actual magnitude of the theoretical uncertainty in $C_\ell$
remains uncertain.

A second explanation relates to the fact that the probability density
function for the field to field distribution of $C_\ell$, P($C_\ell$),
is inherently non-Gaussian, with a long tail towards high values of
$C_\ell$. The significant non-Gaussian contribution is due to Poisson
fluctuations in the sampling of the halo mass function for a given
survey field. Hence, although the average power $\bar C_\ell$ measured
over many maps should reproduce the predicted halo model spectrum
(assuming the cluster gas physics is correctly modeled), the power
measured in any single map may be several times greater than the
ensemble-averaged value, if that field happens to contain several very
massive clusters. These objects, while not contributing strongly to
the mean SZ power spectrum drive the significant non-Gaussian
contribution to P($C_\ell$). \citet{Zhang:07} derived an analytic
expression for P($C_\ell$) using the halo model approach and
demonstrated that the degree of non-Gaussianity is most significant at
large angular scales, where massive clusters contribute most of their
signal. Until now, it has not been possible to generate a large enough
sample of independent simulated SZ maps in order to test the analytic
expressions.

The main purpose of this paper is to demonstrate, using both halo
model calculations and a large sample of simulated SZ sky maps
generated from a `lightcone' simulation, that the non-Gaussianities in
P($C_\ell$) are indeed driven by shot noise due to the number of very
high mass clusters in a given field, and that one can reduce this
Poisson contribution to the total variance in $C_\ell$ by masking
these objects from sky maps. We compare the power spectra measured
from our simulated maps with the halo model analytic predictions for
$\bar C_\ell$ and the non-Gaussian contribution to P($C_\ell$). In
Section \ref{sec:halomodel} we describe the halo approach to
calculating the SZ angular power spectrum and trispectrum, and
demonstrate how masking clusters will increase the sensitivity of the
measured $C_\ell$ to $\sigma_8$. In Section \ref{sec:methods} we
compare the analytic and simulated results for $C_\ell$, P($C_\ell$)
and $\sigma_{\rm NG}$, and demonstrate in practice the utility of
masking the brightest clusters. In Section \ref{sec:foregrounds} we
discuss the impact of astrophysical foregrounds and instrument noise
on our results.

Henceforth, in this paper cluster mass is measured within the
spherical region $R_{vir}$, defined as the region enclosing a mean
overdensity of $\Delta_{\rm vir}\rho_{crit}$. $\Delta_{\rm vir}$ is
calculated using the fitting formula provided by \citet{Bryan:98}. The
fiducial assumed cosmology has parameters $\Omega_m=0.27$, $n_s=0.96$,
$\Omega_\Lambda=0.73$, $\sigma_8=0.77$ and $h=0.72$.

\section{Sunyaev-Zel'dovich Angular Power Spectrum}
\label{sec:halomodel}

The thermal SZ effect is a distortion of the CMB caused by inverse Compton
scattering of CMB photons (at temperature $T_{\rm CMB}$) off electrons (at
$T_e$) in the high temperature plasma within galaxy clusters. To first
order, the temperature change at frequency $\nu$ of the CMB is given
by $\Delta T_{\nu}/T_{\rm CMB} = f_{\nu}(x) y$, where $f_{\nu}(x) =
x(\coth(x/2) - 4)$, $x = h\nu / k_B T_{\rm CMB}$, and $y$ is the normal
Compton parameter
\begin{equation}
y = \left(\frac{k_B \sigma_T}{m_e c^2}\right)\int n_e(l)T_e(l) dl \;,
\end{equation}
where the integral is along the line of sight. There is a host of
higher order effects that can be important at the $10\%$ level,
including bulk motions and relativistic effects
\citep{Nozawa:06}. Unless stated otherwise we assume henceforth that
$f_{\nu}(x) = -2$, which is appropriate in the Rayleigh-Jeans limit.

We can also calculate the SZ power spectrum, assuming that clusters
are Poisson-distributed, by simply summing up the squared
Fourier-space SZ profiles, $\tilde{y}(M,z,\ell)$ of all clusters:
\begin{equation}
C_{\ell} =  \int dz {dV \over dz } \int d \ln M {dn(M,z) \over d \ln M}
\tilde{y}(M,z,\ell)^2
\end{equation}
where V(z) is the comoving volume per steradian and $n(M,z)$ is the
number density of objects of mass $M$ at redshift $z$. For the latter
we use the fitting function of \citet{Jenkins:01}, while for the SZ
profiles as a function of mass and redshift we use either a simple
$\beta$-model (where $y = y_0(1+(\theta/\theta_c)^2)^{-1}$) with
self-similar scaling for the central $y_0$ \citep{Holder:99} or the
hydrostatic model of \citet{Komatsu:02} (henceforth referred to as the
KS model). We have extensively experimented with different profiles,
with most prescriptions showing differences from each other at the
level of tens of percent in the power for consistent assumptions about
gas fractions and cosmological parameters.

The dominant contribution to the non-Gaussian error comes from the
shot noise in galaxy clusters. This Poisson contribution to the
variance at a single value of $\ell$ comes from the one-halo
contribution to the trispectrum \citep{Cooray:01,
Komatsu:02,Zhang:07}. For closely separated bins in $\ell$ space, the
variance in the SZ power spectrum is given by

\begin{equation} 
\sigma^2({C_\ell}) = f_{sky}^{-1} \Bigl [ {2 C_\ell^2 \over (2 \ell +1)\Delta \ell}
+ {T_{\ell \ell} \over 4 \pi} \Bigr ]
\label{eqn:variance}
\end{equation}
where the trispectrum due to shot noise is 
\begin{equation}
T_{\ell \ell} =  \int dz {dV \over dz } \int d \ln M {dn \over d \ln M}
\tilde{y}(M,z,\ell)^4
\end{equation}
Henceforth, we will refer to the second term within the brackets in Eqn.
\ref{eqn:variance}, as the `non-Gaussian' variance (or trispectrum),
$\sigma^2_{\rm nG}$. We will refer to the first term as the `Gaussian'
variance, $\sigma^2_{\rm G}$. We note that the ratio of these two
quantities is independent of the observed map area $f_{sky}$.

\begin{figure}
\plotone{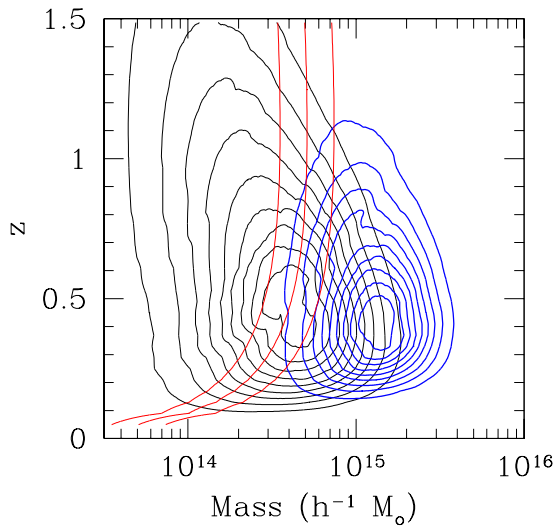}
\caption{Contour plot demonstrating the weight with which clusters of
mass $M_{\rm vir}$ at redshift $z$ contribute to the SZ power spectrum
(black contours) and trispectrum (blue contours) at $\ell = 2500$. The
three lines intersecting the contours are lines of constant integrated
SZ flux (Y), each separated by a factor of 2. They demonstrate that
one could mask the signal from clusters brighter that some threshold
in $Y$ to significantly reduce the trispectrum variance without
removing a large fraction of the signal from the power spectrum.}
\label{fig:cl_mz}
\end{figure}

Figure \ref{fig:cl_mz} demonstrates how clusters at a given mass and
redshift contribute to $C_\ell$ and $T_{\ell \ell}$ at $\ell = 2500$
and for $\sigma_8 = 0.77$ (assuming a $\beta$-model for the gas
pressure profiles). The black and blue contours represent equal
contribution to $C_\ell$ and $T_{\ell \ell}$, respectively. Clusters
in the redshift range $0.3 < z < 0.6$ and of mass $4\times10^{14}\,
h^{-1} M_\odot$ contribute most to the power spectrum, whereas the
trispectrum is driven by clusters at approximately the same redshift
but with more than twice the mass. Since the SZ flux scales as
$(mass)^{5/3}$, the trispectrum is clearly skewed toward much larger
masses than the power spectrum. This leads to a significant amount of
non-Gaussianity in the statistics of the SZ power spectrum being
driven by the most massive objects.

It is clear from Figure \ref{fig:cl_mz} that although the most massive
clusters do not strongly contribute to the mean power spectrum at
$\ell = 2500$, they dominate the trispectrum. These clusters are a
significant source of noise in any measurement of $C_\ell$. Therefore,
if one simply clips the most massive objects from the map, the
field-to-field variance of $C_\ell$ will be reduced and the statistics
will become more Gaussian. The three lines intersecting the contours
in Figure \ref{fig:cl_mz} are lines of constant integrated flux (Y),
each separated by a factor of 2. They represent thresholds above which
one could mask clusters so that the impact of the trispectrum (or
non-Gaussian variance) is greatly suppressed, while not removing a
significant amount of signal from a power spectrum measurement. This
then allows for a more precise measurement of the `true' or
ensemble-averaged SZ power spectrum amplitude, $\bar C_\ell$, and thus
on $\sigma_8$.

\begin{figure}
\plotone{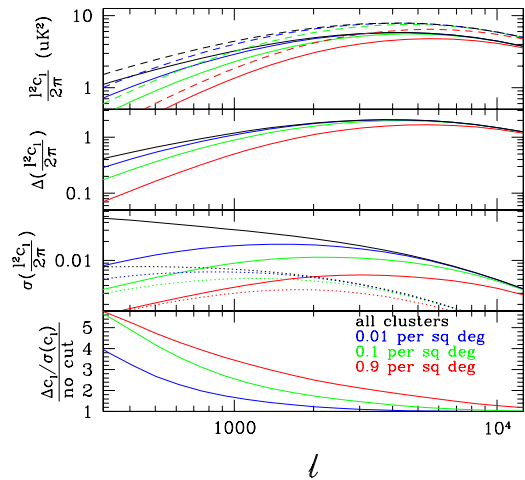}
\caption{({\it top}) The SZ angular power spectrum without masking
  clusters (black), and masking out the brightest 0.01, 0.1 and 0.9
  clusters per square degree (blue, green and red lines). The solid
  and dashed lines show the results for $\sigma_8 = 0.77$ and $0.80$,
  respectively. ({\it middle, top}) The difference in
  $\ell(\ell+1)C_\ell/2\pi$ between the two values of $\sigma_8$ for
  each (un)masked case. ({\it middle, bottom}) The standard deviation
  of the canonical distribution of $C_\ell$, as given by
  Eqn. \ref{eqn:variance}. Note that the value of $\sigma$ plotted
  here is calculated assuming full sky coverage ($f_{\rm sky} = 1$)
  and for $\sigma_8 = 0.77$. The dotted lines shows the contribution
  of the purely Gaussian component ($\sigma_{\rm G}$, see
  Eqn. \ref{eqn:variance}). ({\it bottom}) The sensitivity of the SZ
  power spectrum to a change of 0.03 in $\sigma_8$, $\Delta
  c_\ell/\sigma(C_\ell)$, having masked out clusters divided by the
  sensitivity for the unmasked case. This panel demonstrates that
  masking the brightest clusters down to a surface density of 0.9
  clusters deg$^{-2}$ improves the sensitivity of the SZ power
  spectrum to $\sigma_8$ by a factor of 3.2 (2.0) at $\ell = 1000$
  ($4000$).}
\label{fig:dcldsig8}
\end{figure}

Figure \ref{fig:dcldsig8} demonstrates the improvement in sensitivity
to $\sigma_8$ when clusters above a given threshold are masked out of
the analysis.  We define our threshold in terms of a cumulative
surface density on the sky, such that all bright clusters down to
surface density $n_{\rm th}$ (deg$^{-2}$) are masked out from the
map. We adopt a surface density threshold rather than an integrated
flux threshold so that the number of clusters masked is relatively
insensitive to $\sigma_8$ and modeling of cluster gas physics (see
discussion below). In practice we have chosen three values of $n_{\rm
th}$: 0.01, 0.1 and 0.9 deg$^{-2}$. SPT has already demonstrated that
clusters can be detected at $>5\sigma$ significance down to the 0.1
deg$^{-2}$ threshold \citep{Staniszewski:08}, and both ACT and SPT are
predicted to detect clusters down to the 0.9 deg$^{-2}$
threshold. Therefore, it should be possible for both these experiments
to identify and mask or remove clusters down to the thresholds that we
examine here.

The top panel in Figure \ref{fig:dcldsig8} shows the SZ angular power
spectrum for no masking (black) and masking down to 0.01, 0.1 and 0.9
clusters per square degree (blue, green and red lines). The solid and
dashed lines show the results for $\sigma_8 = 0.77$ and $0.80$,
respectively. As expected, masking the brightest clusters primarily
removes power at larger scales ($\ell<3000$). Removing the brightest
0.01 clusters per square degree has little impact above $\ell=1000$,
however masking down to 0.1 (0.9) deg$^{-2}$ reduces the power
spectrum at this scale by 33\% (66\%). It is clear that the change in
power due to masking is greater than that due to the 0.03 change in
$\sigma_8$, so the masking must be included in the theory prediction
for the signal.

Although the second panel indicates that an unmasked map has the
greatest raw sensitivity to $\sigma_8$, the third panel demonstrates
that the masked maps have much lower total variance, and are therefore
more precise. The reduced variance more than compensates for the
slight reduction in raw sensitivity, such that the ability to use the
SZ power spectrum as a cosmological tool is enhanced by using maps
with the brightest sources masked. This does not include the extra
information encoded in these individually-detected clusters and is
based solely on the information in the power spectrum. The bottom
panel shows the improvement in true sensitivity ($\Delta (C_\ell) /
\sigma(C_\ell)$) having cut out bright clusters, relative to the uncut
case. At $\ell=1000$, masking 0.1 clusters/deg$^2$ results in a factor
of 2.6 improvement in the cosmological sensitivity of the SZ power
spectrum. At $\ell=3000$ there is still a 50\% improvement. 

We note that given full knowledge of the number density of clusters
and the thermal properties of the ICM over a wide range in mass and
redshift, one could construct an optimized scheme for measuring the
power spectrum amplitude by de-weighting the signal from high mass
clusters. However, as one would already know the cluster mass function
-- which is more sensitivity to cosmological parameters -- the power
spectrum would then contain no further cosmological information. In
the absence of such an optimal estimator, a simple solution is to
apply zero weight (i.e. mask) the highest mass clusters from power
spectrum measurements (to which they mostly contribute just noise)
while utilizing them to sample the cluster mass function, to which
they contribute significant statistical information.

\begin{figure}
\plotone{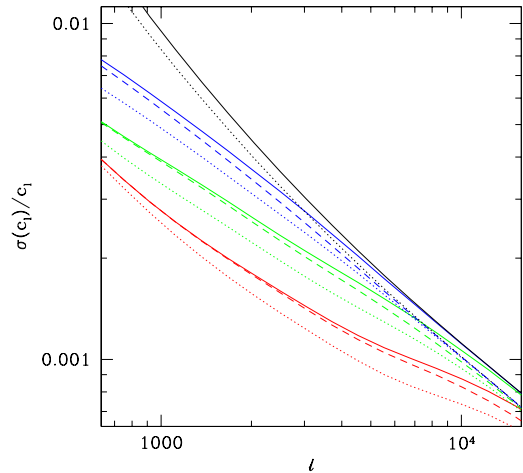}
\caption{The fractional error in the amplitude of the SZ power
spectrum as a function of angular multipole number (assuming full sky
coverage). Solid lines show the results for $\sigma_8 = 0.77$, having
applied no masking (black), and masking the brightest clusters down to
a surface density $n_{\rm th}$ = 0.01 (blue), 0.1 (green) and 0.9
(red) clusters deg$^{-2}$. Dotted lines show the fractional error for
$\sigma_8 = 0.87$ having applied the same cut in flux, $Y_{\rm
th}(n=n_{\rm th})$. Dashed lines show the results for $\sigma_8 =
0.87$ having applied surface density cuts rather than flux cuts.}
\label{fig:densitycut_fluxcut}
\end{figure}

We now investigate the difference between applying cuts defined by a
surface density or integrated SZ flux threshold. In Figure
\ref{fig:densitycut_fluxcut} we plot the fractional error in the
amplitude of the SZ power spectrum as a function of angular multipole
number (assuming full sky coverage). The black solid and dotted lines
show the unmasked results for $\sigma_8 = 0.77$ and $0.87$
respectively. The solid blue, green and red curves show the results
having applied our normal surface density cuts ($n_{\rm th}$ = 0.01,
0.1 and 0.9 clusters deg$^{-2}$) for $\sigma_8 = 0.77$.  The dotted
lines show the fractional error for $\sigma_8 = 0.87$ having applied
exactly the same cut in flux. The dashed lines show the results for
$\sigma_8 = 0.87$ having now used the surface density rather than flux
thresholds.

The plot demonstrates that using cuts in surface density produces a
fractional error that is approximately independent of $\sigma_8$. On
the other hand, adopting a flux threshold means that the number of
clusters masked -- and thus $\sigma(C_\ell)/C_\ell$ -- is highly
sensitive to $\sigma_8$. Larger values of $\sigma_8$ result in greater
numbers of clusters with flux greater than a given value. This will
ultimately make the statistical error on measurements of $C_\ell$ more
difficult to determine. Furthermore, the effectiveness of applying
flux cuts is also dependent on the accuracy with which one is able to
measure cluster fluxes observationally, and the mass and redshift
dependence of the intrinsic scatter in the M-Y relation. Using surface
density thresholds neatly circumvents these issues.

\section{Comparisons with Simulations}
\label{sec:methods}

\subsection{Simulations}

To date, it has proven difficult to compare the statistics of the SZ
power spectrum as predicted by analytic calculations and numerical
simulations as the typical box size of hydrodynamical simulations
limits the number of independent maps that can be created, while also
poorly sampling the extreme high-mass end of the halo mass
function. Here we combine an N-body `lightcone' simulation with a
semi-analytic model for cluster gas to generate a large number of maps
and thus enable such a comparison. We also use our large sample of
simulated maps to demonstrate the utility of masking clusters to
improve the precision of measurements of the SZ power spectrum.

To generate the simulated SZ maps, we begin with the output of a large
($N=1260^3=2\times 10^9$ particles) cosmological dark matter
simulation. Cosmological parameters were chosen to be consistent with
those measured from the 3rd-year WMAP data combined with large-scale
structure observations \citep{Spergel:07}. The simulated volume is a
periodic cube of size $L=1500h^{-1}$Mpc; the particle mass
$m_p=1.22\times 10^{11}h^{-1}M_\odot$ and the cubic spline softening
length $\epsilon=16.5h^{-1}$kpc. This simulation was carried out with
exactly the same methods used for the simulation described in
\citet{Bode:07}, \citet{Sehgal:07} and \citet{Shaw:08}, so for a
description of the computational details we refer the reader to these
papers.

The very large box size of this simulation makes it ideal for
generating a large area of simulated sky. The matter distribution in a
light cone extending to $z=3$ was saved in 348 time slices. Dark
matter halos in this light cone were identified with the
Friends-of-Friends algorithm using a comoving linking length parameter
$b=0.2$. In total, the lightcone covers a single octant on the sky
($\approx {\rm 5000} $ deg$^2$), containing over two million clusters
with $M_{\rm FOF} > 10^{13} h^{-1} \msun$.  For the purposes of the
analysis here we retain only clusters with $M_{\rm FOF} >
3\times10^{13} h^{-1} \msun$.

The cluster gas distribution in each halo was calculated using the
semi-analytic model described in \cite{Ostriker:05} and
\citet{Bode:07}.  In brief, a 3D mesh (with cell side-length
$2\epsilon = 33 h^{-1}$ kpc) is placed around each cluster, with the
gas pressure and density determined in each mesh cell assuming
hydrostatic equilibrium and a polytropic equation of state \citep[with
adiabatic index $\Gamma=1.2$,][]{Ascasibar:06}.  It is also assumed
that the gas in the densest cluster regions has cooled and condensed,
forming stars. At z = 0, the stellar/gas mass ratio is set to 0.1
\citep{Lin:03, Voevodkin:04}. To compute the star/gas ratio at $z >
0$, the star formation rate was assumed to follow a delayed
exponential model \citep[Eqn. 1 of][]{Nagamine:06}, with decay time
$\tau=1.5$ Gyr.

As discussed in detail in \cite{Bode:07}, the most important free
parameter in this model is the energy input into the cluster gas via
non-thermal feedback processes, such as AGN outflows and supernovae.
This is set through the parameter $\epsilon_{f}$, such that the
feedback energy is $\epsilon_{f}M_*c^2$, where $M_*$ is the stellar
mass in the cluster. It was demonstrated that setting $\epsilon_f =
3\times10^{-5}$ provides good agreement between the $M -T$ and $f_g -
T$ scaling relations obtained from our simulation plus gas model
cluster sample and those from X-ray observations \citep{Arnaud:05,
Vikhlinin:06, Gastaldello:07, oHara:06, Zhang:07b}.

From the output of our ICM model, we produce a 2-d `postage-stamp' $y$
image for each cluster by summing up the electron pressure $n_e T_e$
in each mesh cell along one direction \citep[for example, see Fig. 7
of][]{Ostriker:05}. We then construct a library of images for all
clusters in the lightcone octant with $M_{\rm vir} > 2.5 \times
10^{13} \msun$. Note that each postage-stamp is produced having
rotated the 3D mesh around the cluster so that one face lies parallel
to the plane of the lightcone `sky'. SZ cluster sky maps are thus
produced by projecting down the lightcone, summing up the contribution
of all the clusters along the line of sight. In practice, we produce
individual SZ sky maps of size 5x5 degrees. We set the image
resolution (pixel size) to 0.25 arcminutes. In order to allow robust
comparison with the analytic results, we also create a second set of
simulated maps in which we replace the gas model cluster
`postage-stamp' images with a $\beta$-model profile with $\beta = 1$
and normalized in the same way as for the analytic calculation
described in the previous Section. We henceforth refer to the maps
produced with the \citet{Bode:07} gas prescription as the BO gas model
sample. In total we produce 96 25 deg$^2$ maps (2400 deg$^2$ in total
of simulated sky) for each gas profile.

\subsection{Simulated SZ power spectrum and trispectrum}
\label{sec:results}

\begin{figure}
\plotone{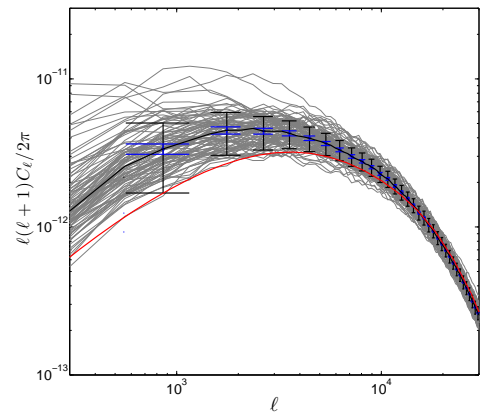}
\caption{SZ angular power spectrum measured for each of the 25 deg$^2$
simulated maps for our BO sample (grey lines). Black error bars plot
the $1\sigma$ statistical errors around the mean value $\bar C_\ell$
(black solid line) for the sample.  The blue error-bars denote the
standard deviation expected from Gaussian cosmic variance alone (the
first term in Eqn. \ref{eqn:variance}). The red line denotes the power
spectra calculated using the halo model and KS gas profile. }
\label{fig:power_allfields}
\end{figure}

In Figure \ref{fig:power_allfields} we plot the SZ angular power
spectrum measured for each of the 25 deg$^2$ simulated maps (grey
lines) in our BO gas model sample. The solid black line represents the
mean power spectrum measured for the BO gas model maps; the red solid
line the halo model predictions for the KS gas model. The black error
bars denote the $1\sigma$ statistical errors around the mean value
$\bar C_\ell$ for the BO map sample.  The blue error-bars denote the
standard deviation expected from Gaussian cosmic variance alone
($\sigma_{\rm G}$, the first term in Eqn. \ref{eqn:variance}).  The
simulated power spectra are band-averaged in bins of $\Delta_l = 300$
and are apodized with a 2d Gaussian window function with unity
variance. The analytic power spectra have been convolved with a
Gaussian of FWHM 0.25 arcminutes in order to approximately account for
the pixelisation of the simulated maps. We also note that the analytic
power spectra are calculated omitting the contribution of clusters
with $M_{\rm vir} < 2.5\times 10^{13} \msun$.  This is because the
simulated maps are only 100\% complete down to this mass, and so we
have omitted lower mass clusters from the simulated sky-maps.

We note that the power spectra measured for the BO gas model
simulations agree with the KS power spectra at very small angular
scales ($\ell>10,000$) but have significantly more power at larger
scales. This additional power is most likely due to the impact of the
feedback parameter in the BO model. The additional energy added to the
gas has the effect of heating and `puffing out' the gas, reducing the
slope of the pressure profiles in the outer regions and increasing the
large-scale signal. We also note that the KS model has a relatively
steep SZ profile at larger radii, reducing the signal at larger
scales. The full impact of non-gravitational heating sources on the
ICM gas temperature and density profiles is currently poorly
understood; the KS presciption is not astrophysically well-motivated
and there is no reason to expect the model to match either the BO gas
model or real clusters at high redshift. Hence there remains a large
theoretical uncertainty in $C_\ell$ which must be accounted for when
attempting to constrain $\sigma_8$ via the SZ power spectrum. The
exact range of uncertainty in $C_\ell$ between different models and
simulations has not been quantified and we leave this issue to future
work. We have confirmed that the power spectra measured from our
sample of $\beta$-model maps matches the equivalent predictions from
the halo model approach.

\begin{figure}
\plotone{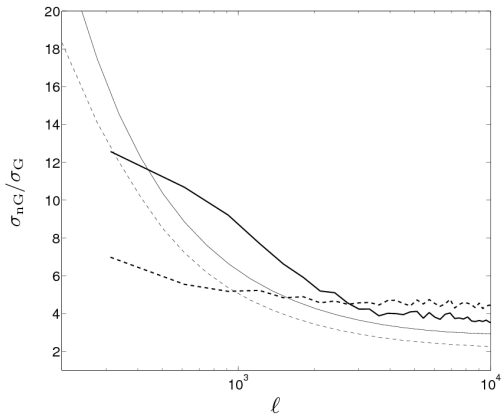}
\caption{Square root of the ratio of the non-Gaussian (trispectrum) to
Gaussian contributions to the total variance in $C_\ell$, $\sigma_{\rm
nG}/\sigma_{\rm G}$, as a fuction of angular multipole number. Heavy
solid and dashed lines represent the $\beta$-model and BO gas model
simulated maps, respectively. Light solid and dashed lines show the
results from the halo-model calculation assuming $\beta$-model and KS
gas profiles.}
\label{fig:tll_allmods}
\end{figure}

As demonstrated by the error bars in Figure \ref{fig:power_allfields},
the total field-to-field variation in the power spectrum amplitude is
several times greater than the Gaussian error, with the factor
decreasing with increasing $\ell$. This is due to Poisson noise
associated with the discrete sampling of the halo mass function in
each field. Comparing the individual simulated power-spectra to the
error-bars, it is also clear that the distribution of $C_\ell$ is
highly non-gaussian, especially for $\ell \leq 6000$.  The power
spectra for individual maps (grey lines) indicate that the
non-gaussian variance is being driven by fields with $C_\ell >> \bar
C_\ell$ (where $\bar C_\ell$ represents the mean), corresponding to
fields that contain at least one very massive cluster.

In Figure \ref{fig:tll_allmods} we plot the ratio of the non-Gaussian
and Gaussian contributions to the total standard deviation in the
statistical distribution of $C_\ell$, $\sigma_{\rm nG}/\sigma_{\rm G}$
(the second term divided by first term in Eqn. \ref{eqn:variance}), as
a function of angular multipole number. The total variance of
P($C_\ell$) is therefore just the sum of $\sigma_{\rm nG}^2$ and
$\sigma_{\rm G}^2$, although we reiterate P($C_\ell$) is a strongly
non-gaussian function at larger scales and thus the total variance
(and mean) does not completely define this distribution.  $\sigma_{\rm
nG}^2$ is measured from the simulated maps by calculating the variance
in $C_\ell$ for each set of maps, multiplying this by $f_{\rm sky}$
and deducting the Gaussian variance, calculated using $\bar
C_\ell$. The heavy lines represent the results for our simulated
$\beta$-model (solid) and BO gas model (dashed) maps. The thin lines
are the equivalent results for the $\beta$-model and KS halo model
calculations, respectively.

Comparing thick and thin solid black lines, we find that the
non-Gaussian error measured for our set of simulated $\beta$-model
maps is approximately one third greater than that predicted by the
trispectrum. As we find a good agreement between the simulated and
predicted power spectra, this discrepency must be due to effects that
increase the width of P($C_\ell$) without influencing strongly the
ensemble-averaged mean. One such effect is the non-linear bias, or
small scale halo clustering, which is not accounted for in the
trispectrum. As noted in Section \ref{sec:halomodel} and
\citet{Cooray:01, Komatsu:02, Zhang:07}, the two halo (and
higher-order) contribution to the trispectrum -- which accounts for
the linear bias -- is insignificant, including the two-halo term
increases $\sigma_{\rm nG}$ by only a few percent. We have found that
by remeasuring $\sigma_{\rm nG}$ for both sets of simulated maps
having first randomised the halo positions reduces $\sigma_{\rm
nG}/\sigma_G$ such that the simulated $\beta$-model line becomes into
better agreement with that of the trispectrum result, although remains
above the predicted curve at all $\ell$.

The thick and thin dashed lines in Figure \ref{fig:tll_allmods} show
the ratio of the non-Gaussian to Gaussian variance for BO gas model
simulations and the analytic KS model. The simulations demonstrate a
shallower dependence on $\ell$ than the halo model calculation. Given
that $\sigma_{\rm G} \propto C_\ell$, the flatness of the simulated
curve is likely due to the larger mean power simulated spectrum in the
simulated maps compared to the KS model predictions. We note again
however that there is no reason why an agreement is expected. Indeed,
as the simulation results include large variations in the underlying
dark matter halo structure (including morphology and substructure) we
would expect $\sigma_{\rm nG}/\sigma_{\rm G}$ to be greater,
especially at small angular scales.

\subsection{Masking Clusters}
\label{sec:masking}

\begin{figure}
\plotone{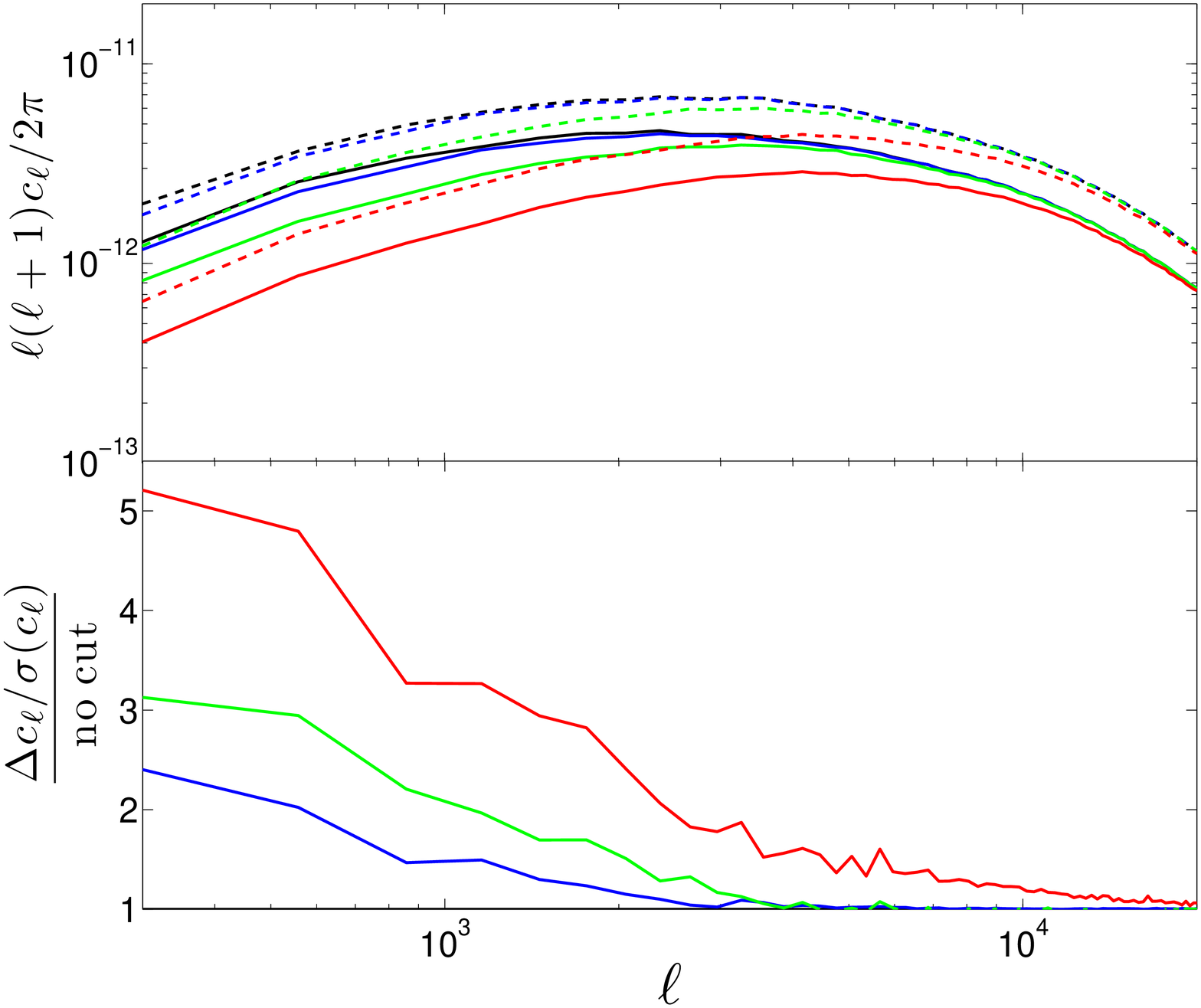}
\caption{({\it Upper}) The mean angular power spectrum measured for
our WMAP 3 (solid) and 5 (dashed) samples of simulated maps (BO gas
model), having applied no masking (black), and masking the brightest
clusters down to 0.01, 0.1 and 0.9 clusters deg$^{-2}$ (blue, green
and red lines). ({\it Lower}) The ratio of the sensitivity of the SZ
power spectrum to the difference in $\sigma_8$ between our two
simulations having masked clusters to that with no masking.}
\label{fig:cl_tll_masked}
\end{figure}

In Section \ref{sec:halomodel} we showed that the primary contribution
to the SZ power spectrum comes from clusters occupying a different
region of the cluster mass-redshift plane to those that determine the
trispectrum.  Clusters at $z \approx 0.4$ and of mass
$4\times10^{14}h^{-1}\msun$ contribute most strongly to the power
spectrum, whereas clusters at $1-2\times10^{15}h^{-1}\msun$ are the
foremost contributors to the trispectrum. Therefore, if one wishes to
measure accurately the mean SZ power spectrum, very high mass clusters
constitute a source of noise. One can mask or remove the brightest
clusters detected in a SZ map down to some threshold density in order
to suppress the trispectrum contribution to the total variance in
$C_\ell$ without removing a large component of the signal from the
power spectrum. We now demonstrate this in practice using our sample
of simulated maps. 

To evalulate the sensitivity of the SZ power spectrum to $\sigma_8$,
we use a second lightcone simulation to create an additional sample of
mock SZ maps. The cosmological parameters were chosen to be consistent
with those measured from the 5th-year WMAP data \citep{Dunkley:09},
therefore $\sigma_8 = 0.8$ for this simulation (compared to $0.77$ as
for our WMAP3 run). The simulated volume is a periodic cube of size
$L=1000h^{-1}$Mpc; the particle mass $m_p=6.82\times
10^{10}h^{-1}M_\odot$ and the cubic spline softening length
$\epsilon=16.28h^{-1}$kpc. We have applied the \citet{Bode:07} gas
prescription to the halos in the WMAP5 lightcone, and created a second
set of mock SZ sky maps using the procedure described in the previous
section. We refer to the $\sigma_8 = 0.77$ maps as the WMAP3
simulation, and the $\sigma_8 = 0.80$ maps as the WMAP5 simulation.

\begin{figure*}
\plotone{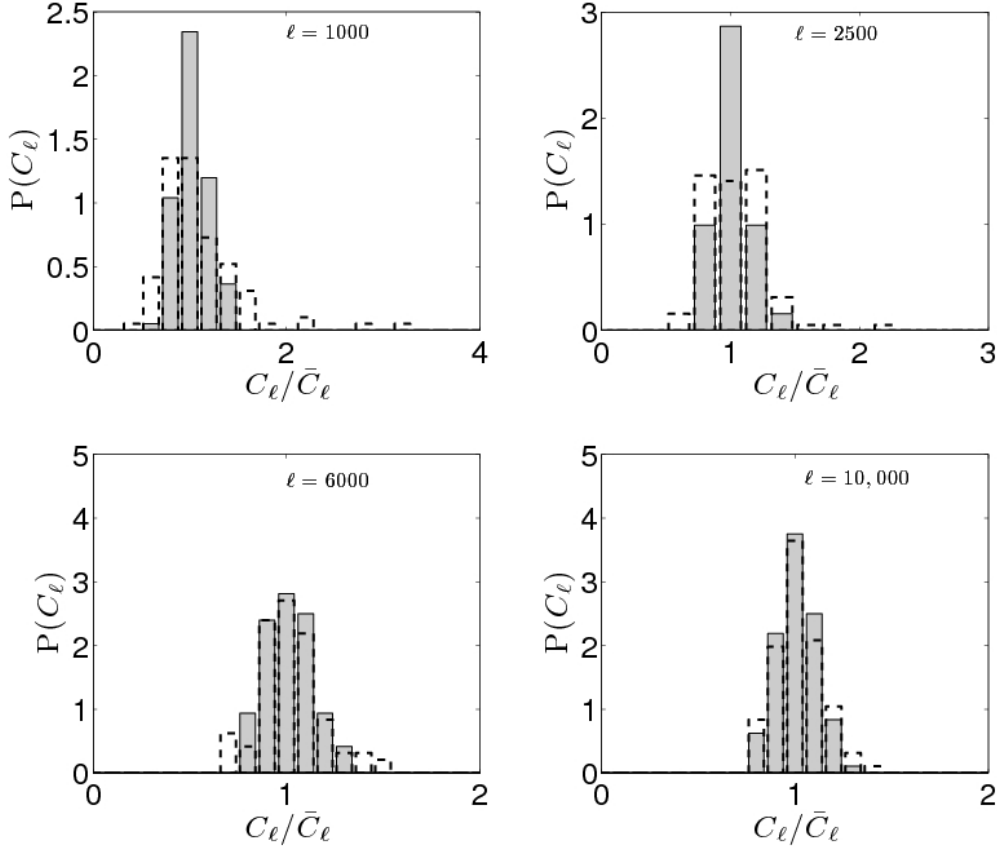}
\caption{The distribution of $C_\ell/\bar C_\ell$ measured for each
simulated BO gas model field in our WMAP3 sample, where $\bar C_\ell$
is the mean over the whole sample of 96 maps, at four different
angular multipole numbers: $\ell = 1000, 2500, 6000$ and $10,000$. The
unshaded, dashed histogram shows the results without masking
clusters. The shaded histogram represents results having masked the
brightest 0.9 clusters/deg$^2$ from the maps.}
\label{fig:pcl}
\end{figure*}

Clusters are masked from the simulated maps in the following
manner. For each of the 96 maps in the WMAP3 \& 5 samples, there is an
associated catalogue listing all clusters in map. We combine all 96
catalogues and rank-order clusters by their integrated SZ flux
$Y$. Starting with the brightest cluster, we then fit and remove an
isothermal $\beta$-model (with $\beta$ fixed at 1) from the map at the
location of the cluster. This is then repeated until all clusters down
to the desired surface density threshold $n_{\rm th}$ have been
masked. This process can be performed in practice on real SZ survey
maps by ranking detected clusters by their detection significance, and
removing each in turn down to the required threshold surface density
\citep[and is somewhat analogous to the application of an optimal
matched-filter, ][]{Herranz:02, Melin:06, Schafer:06}.

The upper panel of Figure \ref{fig:cl_tll_masked} demonstrates the
mean power spectrum measured with no clusters masked (black line), and
having masked clusters down to 0.01, 0.1 and 0.9 deg$^{-2}$ (blue,
green and red lines). The solid lines show the results for our WMAP3
simulation, the dashed lines for our WMAP5 lightcone simulation. For
minimal masking, power is removed only on large angular scales,
shifting to smaller scales as increasing numbers of clusters are
masked. Removing 0.01 clusters per square degree barely changes the
average power spectrum, whilst masking 0.1 clusters per square degree
reduces power by less than a third at $\ell=1000$, and negligibly at
$\ell>3000$. Masking 0.9 clusters deg$^{-2}$ removes power over a wide
range of scales, reducing the mean band-averaged power by a factor of
2.5 at $\ell=1000$ and by 27\% at $\ell=5000$. As expected, the most
massive clusters contribute most of their power at large angular
scales, and that at scales of $\ell>5000$ low mass clusters contribute
the majority of the signal. These results are similar to those shown
in Figure \ref{fig:dcldsig8}.

In the lower panel of Figure \ref{fig:cl_tll_masked} we plot the
sensitivity of the SZ power spectrum to the difference in $\sigma_8$
between the two simulation samples ($\Delta \sigma_8 = 0.03$) including
cluster masking, divided by the sensitivity for the unmasked case. We
define sensitivity as being the difference in $\bar C_\ell$ between the two
simulations divided by the standard deviation in $C_\ell$,
$\sigma(C_\ell)$ measured in the $\sigma_8 = 0.77$ sample of maps. The
line colours are the same as in the upper panel. It is immediately
evident that whilst removing down to 0.01 clusters per square degree
has little impact on the mean power spectrum, there is a clear gain in
sensitivity below $\ell = 1000$. Masking just the 10 brightest
clusters per 100 deg$^2$ improves the sensitivity by a factor of two at
$\ell = 1000$ and 1.5 at $\ell = 2000$. Masking the 90 brightest
clusters in the same area enables similar improvements at $\ell=2000$
and $4000$.  The improvement in sensitivity measured for our simulated
maps agrees well with the analytic predictions from the halo model
approach (see the lowest panel in Figure \ref{fig:dcldsig8}).

In addition, as one masks out the brightest clusters the PDF of
$C_\ell$ becomes significantly more Gaussian. This is demonstrated in
Figure \ref{fig:pcl}, in which we plot the distribution of
$C_\ell/\bar C_\ell$ at l = 1000, 2500, 6000 and 10,000 for our WMAP3
sample. The unshaded, dashed histogram represents P($C_\ell/\bar
C_\ell$) for the unmasked case, the shaded histogram represents
P($C_\ell/\bar C_\ell$) having masked bright clusters down to 0.9
clusters per square degree. For the unmasked case, it is clear that
the level of non-Gaussianity is significantly greater at large scales,
where massive clusters contribute most of their power. It is also
immediately apparent that the masked distribution is significantly
more Gaussian than the unmasked case. This can be quantified by the
reduction in skewness between the masked and masked distributions,
which is 1.0, 0.5, 0.2 and 0.1 at $\ell$ = 1000, 2500, 6000 and
10,000.

\section{Impact of Foregrounds and Noise}
\label{sec:foregrounds}

We have demonstrated the utility of masking clusters from SZ sky maps
to reduce the non-Gaussian variance in the amplitude of the SZ power
spectrum and thus improve the precision with which one can measure the
ensemble-averaged $C_\ell$. However, there still remains a question of
the overall gain in sensitivity once one accounts for astrophysical
foregrounds, such as the primary CMB temperature anisotropies and the
contribution from radio and sub-mm point source populations, and
instrument noise. While masking bright clusters reduces the
field-to-field variance in $C_\ell$, the amplitude of the
ensemble-averaged $C_\ell$ is also reduced. Masking increasing numbers
of clusters thus reduces the amount of SZ power relative to that of
the foregrounds. The variance in the foreground signal is proportional
to $C_{\ell,{\rm fg}}^2$, therefore as the SZ signal is reduced by
masking, the total fractional error on the SZ amplitude due to the
error in the foreground signals increases. The relevant question is,
how does the total intrinsic error (Gaussian plus non-Gaussian) on the
SZ signal $\sigma(C_{\ell,{\rm sz}})$ compare to the error (or
uncertainty) on the the primary CMB and point source power spectra
amplitudes within a given field?

If $\sigma_{\rm nG}$ dominates the error in the total power measured
on the sky, $C_{\ell,{\rm sky}}$, where
\begin{equation}
C_{\ell,{\rm sky}} = C_{\ell,{\rm sz}} + C_{\ell,{\rm CMB}} + C_{\ell,{\rm PS}} + C_{\ell,{\rm inst}}\;,
\end{equation}
then one can simply remove the foreground components from the measured
power spectrum, and clipping the bright clusters will improve the
precision on the measured $C_{\ell,{\rm SZ}}$. On the other hand, if
the error due to cosmic variance on $C_{\ell,{\rm CMB}}$ or
$C_{\ell,{\rm PS}}$ is comparable to $\sigma(C_{\ell,{\rm SZ}})$ then
reducing the amplitude of the SZ signal by masking clusters may
degrade the measured SZ power spectrum amplitude.

For mm-wavelength experiments, the variance on the total power
spectrum due to all sources can be written as the sum in quadrature of
the Gaussian sample variance term (including CMB, point sources,
Gaussian SZ error and instrument noise), the non-Gaussian SZ error and
a theory error,
\begin{equation}
\sigma^2(C_{\ell, {\rm sky}}) = (\sigma_{\rm cmb} + \sigma_{\rm sz,G}
+ \sigma_{\rm ps} + \sigma_{\rm inst})^2 +\sigma^2_{\rm sz,nG} +
\sigma^2_{\rm th}\;,
\end{equation}
where we have included an error on the predicted SZ power spectrum,
$\sigma^2_{\rm th}$, due to the uncertainties in modeling cluster gas
physics. Using Eqn. \ref{eqn:variance} this can be rewritten as
\begin{equation}
\frac{\sigma^2(C_{\ell, {\rm sky}})}{(M C_{\ell,{\rm sz}})^2} =
\frac{2f^{-1}_{\rm sky}}{(2\ell+1)\Delta \ell} \left[(1+F/M)^2 +
s^2\right] + t^2 \;,
\label{eqn:total_variance}
\end{equation}
where $s(\ell,n_{\rm th}) = \sigma_{\rm nG}/\sigma_{\rm G}$ and
$M(\ell,n_{\rm th})$ is the ratio of the masked to unmasked
$C_{\ell,{\rm sz}}$'s for a given mask surface density threshold
$n_{\rm th}$, and $t$ is the fractional error on $C_{\ell,{\rm sz}}$
due to the theoretical uncertainties in cluster gas physics. $F$ is
defined as the ratio of foreground to SZ power,
\begin{equation}
F(\ell) = \frac{C_{\ell,{\rm CMB}} + C_{\ell,{\rm PS}} + B^{-2}C_{\ell,{\rm inst}}}{C_{\ell,{\rm sz}}} \;,
\end{equation}
where $B$ is the Fourier domain profile of the telescope beam. We
henceforth assume a 10\% theoretical uncertainty on $C_\ell$, and
$10\mu K$ instrument noise in an arcminute pixel for a
single-frequency survey at 90Ghz of 25 deg$^2$ of sky with a Gaussian
beam of FWHM 1.67 arcminutes. The values of $s$ are measured from our
WMAP3 simulation SZ map sample, with $\Delta \ell = 300$. The point
source power spectrum is calculated assuming radio and sub-mm source
populations and spectral index distributions as given in
\citet{deZotti:05} and \citet{Negrello:07}.

\begin{figure}
\plotone{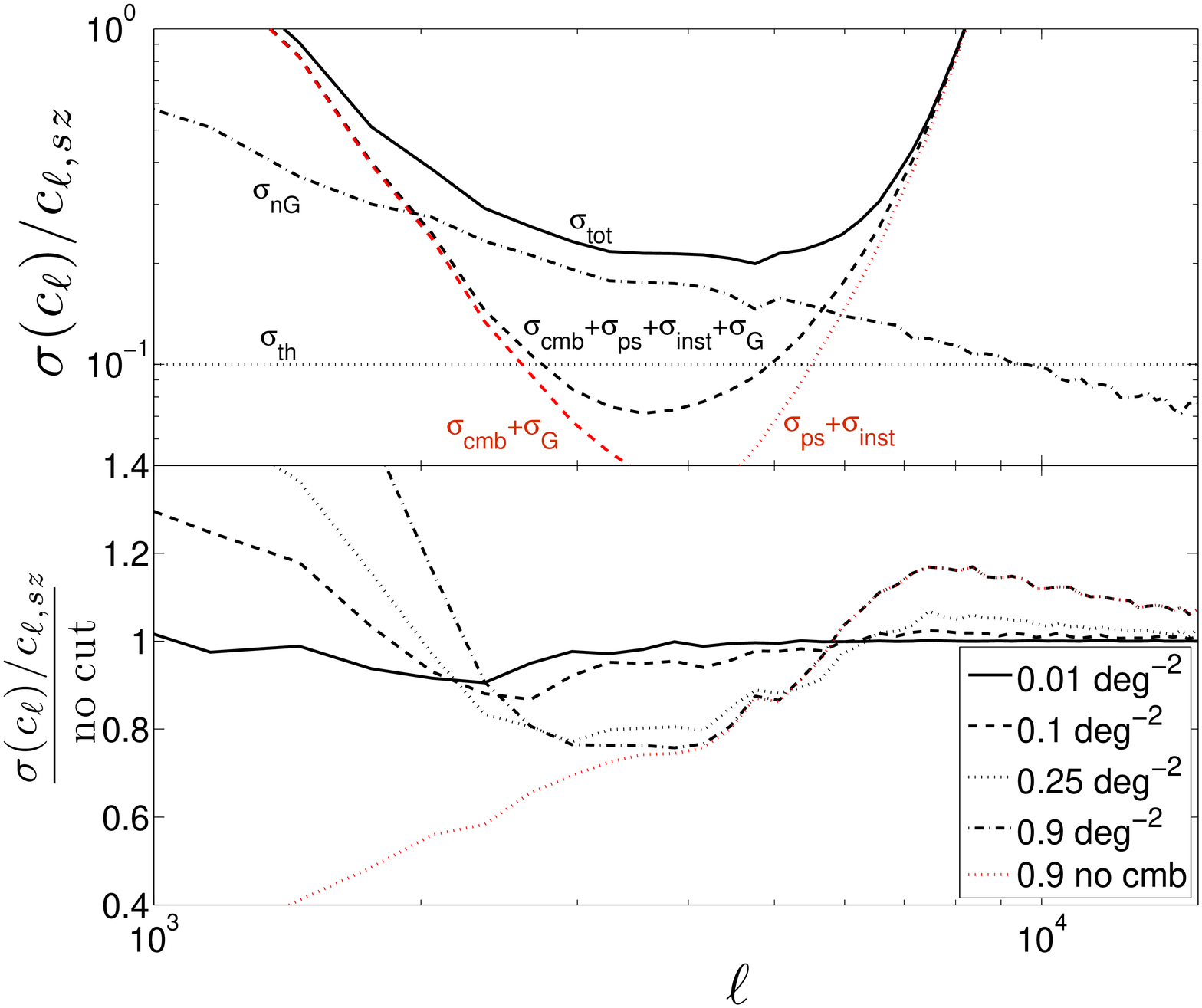}
\caption{({\it Upper}) Fractional error on $C_{\ell,{\rm sz}}$ as a
function of angular multipole number. The solid black line gives the
total fractional error as given in Eqn. \ref{eqn:total_variance}. The
black dot-dashed, dashed and dotted lines show the contributions of
the SZ trispectrum, the combined foreground signals (plus Gaussian SZ
variance), and the theory uncertainty on $C_\ell$, respectively. The
red dashed line shows the contribution of just the primary CMB and
Gaussian SZ variance; the red dotted line shows that of the point
sources plus instrument noise. ({\it Lower}) The total fractional
error on $C_{\ell,{\rm sz}}$ having masked bright clusters from the
map divided by the error on $C_{\ell,{\rm sz}}$ having applied no
masking. In the range $3000 < \ell < 6000$ (where the SZ trispectrum
is the dominant error) removing bright clusters from the map increases
the precision of measurements of the SZ power spectrum.}
\label{fig:frac_var_foregrounds}
\end{figure}

The upper panel of Figure \ref{fig:frac_var_foregrounds} shows the
total fractional error on $C_{\ell,{\rm sz}}$ as a function of angular
multipole number. The solid black line gives the total fractional
error (as given in Eqn. \ref{eqn:total_variance}) for no masking
($M=1$). The black dot-dashed, dashed and dotted lines show the
contribution of the SZ trispectrum, the total error on the combined
foreground signals (including instrument noise), and the theory
uncertainty on $C_\ell$, respectively. The red lines show the
contributions of the CMB plus Gaussian SZ variance (dashed) and that
of the point sources and instrument noise (dotted). It is evident that
in the range $2300 < \ell < 5000$ the non-Gaussian variance in the SZ
power spectrum amplitude is the dominant source of error on
$C_{\ell,{\rm sz}}$, with cosmic variance in the CMB power dominating
at larger scales, and that of the point sources and instrument noise
dominating at smaller scales. It is clear that in this range,
$C_{\ell,{\rm sz}}$ can be more accurately measured if one can reduce
the non-Gaussian variance $\sigma_{\rm nG}$ by masking clusters.

The lower panel of Figure \ref{fig:frac_var_foregrounds} demonstrates
the reduction in the fractional error on $C_{\ell,{\rm sz}}$ having
masked bright clusters down to the surface density threshold $n_{\rm
th}$ (given in the legend), compared to the unmasked case. For $\ell <
2000$, the Gaussian variance on the CMB amplitude dominates and so
masking down to 0.1 clusters deg$^{-2}$ (dashed line) merely reduces
the amplitude of the SZ signal with no associated reduction in the
error ($\sigma(C_\ell)$). Thus the total fractional error increases
relative to the unmasked case. However, masking clusters down to 0.25
per square degree (dotted) gives a 20\% reduction in the fractional
error on $C_{\ell,{\rm sz}}$ compared to the unmasked case in the
range $3000 < \ell < 4000$. We have found that 0.9 clusters deg$^{-2}$
(dash-dotted) gives the best results; more aggressive masking reduces
the amplitude of the SZ power to the point at which the error on the
foregrounds becomes significant. Above $\ell = 6000$, the variance due
to point sources and instrument noise dominates and so masking
clusters simply diminishes the amplitude of the SZ signal without an
associated reduction in the variance. As at $\ell < 2000$, this
results in a decreased sensitivity to $C_{\ell, {\rm sz}}$ at these
scales. The red dotted line shows the reduction in the fractional
error (for $n_{\rm th} = 0.9$) if one can perfectly remove the primary
CMB contribution using multi-frequency observations.

Overall, our results show that, after accounting for the various
foregrounds, for a simulated single frequency survey with $10\mu K$
instrumental noise a 20\% improvement in the measurement of the
amplitude of the SZ angular power spectrum can be obtained if one
masks out the brightest clusters from the map down to a surface
density threshold of at least 0.25 clusters degree$^{-2}$. This
corresponds to a reduction of $\approx 3\%$ on the error on
$\sigma_8$, although we note the additional gain in the Gaussianized
statistics of P($C_{\ell,{\rm sz}}$).

\section{Conclusions}

We have demonstrated that the SZ power spectrum has a significant
amount of variance contributed by rare massive clusters. While these
objects do not contribute heavily to the ensemble-average SZ power
spectrum, they do add significant non-Gaussian noise to measurements
of the power spectrum amplitude and thus degrade cosmological
utility. Furthermore, it is difficult to use numerical simulations to
characterize the statistics of the SZ power spectrum because one must
have sufficiently large volumes to be able to adequately sample the
rarest objects in the Universe. We have also shown that the magnitude
and scale dependence of the non-Gaussian variance is sensitive to the
modelling of the intra-cluster gas.

A simple solution is to mask out the largest objects in the map before
estimating the power spectrum. We have demonstrated using both halo
model calculations and a large sample of simulated SZ sky maps that
fairly conservative masking can reduce the variance and Gaussianize
the statistics significantly, increasing the precision of the measured
$C_\ell$ and thus the sensitivity to cosmological parameters. For
example, masking the 10 brightest clusters in a 100 deg$^2$ SZ map
will improve the precision on $C_\ell$ by a factor of two at $\ell =
1000$ and 1.5 at $\ell = 2000$. Masking the 90 brightest clusters in
the same region should improve the accuracy by the same factors at
$\ell = 2000$ and $4000$. This is in agreement with work related to
information content in the matter power spectrum in the trans-linear
regime, where it has been found that a few massive objects are
responsible for much of the variance
\citep{Neyrinck:08}. \citet{Takada:07} have also demonstrated that
knowledge of the number of massive clusters within a given survey
field can help improve the precision of measurements of the weak
lensing shear power spectrum.

Choosing which objects to mask is non-trivial; we found that using a
fixed sky density produced a well-defined and well-behaved estimate
that can easily be applied to real maps. This skirts the issue (at
least partially) of finding an observable that is well-correlated with
mass, and can be easily applied to any survey (for example, weak
lensing).

We have also shown that the non-Gaussian variance in the SZ power
spectrum remains a significant source of error in the range $2300 <
\ell < 5000$, even in the presence of the astrophysical foregrounds
provided by the primary CMB temperature anisotropies and unresolved
point source populations. We have demonstrated that even single
frequency experiments with no ability to spectrally subtract the CMB
and point sources, can improve the accuracy on $C_{\ell,{\rm sz}}$ by
more than 20\% in the range $3000 < \ell < 4000$ by masking the
brightest clusters down to between 0.25 and 0.9 clusters per square
degree.

We emphasize that one is not losing cosmological information by
masking bright clusters since our procedure is a way to discard the
signal from massive clusters whose information we do not know how to
analyze but which still contributes to the variance. Halo model
calculations of the SZ power spectrum demonstrate that the
ensemble-averaged power spectrum ($\bar C_{\ell,{\rm sz}}$) -- to
which high mass clusters do not contribute significantly -- has a
sensitive dependence on $\sigma_8$ \citep{Komatsu:02}. If one had
prior knowledge of the number density and gas properties of clusters
over a wide mass and redshift range it would be possible to construct
an optimal power spectrum estimator in which the signal from high mass
clusters would be down-weighted. However, in this case one would
already know the cluster mass function -- which is more sensitive to
cosmological parameters -- and thus power spectrum measurements would
reveal no additional information. In the absence of complete knowledge
of the mass function, one can simply apply zero weight to the
brightest clusters as they effectively contribute only noise to the
power spectrum. On the other hand, masked clusters carry much
statistical weight when used to measure the cluster mass function.

There are still theory challenges to predicting the detailed shape and
amplitude of the SZ power spectrum, but we have demonstrated that the
field-to-field variance can be reduced to allow precise tests of the
physics and astrophysics involved.

\section{ACKNOWLEDGMENTS}
This work supported by the Natural Sciences and Engineering Research
Council (Canada) through the Discovery Grant Awards to GPH. GPH would
also like to acknowledge support from the Canadian Institute for
Advanced Research and the Canada Research Chairs Program. This
research was facilitated in part by allocations of time on the COSMOS
supercomputer at DAMTP in Cambridge, a UK-CCC facility supported by
HEFCE and PPARC. Computer simulations were supported by the National
Science Foundation through TeraGrid resources provided by Pittsburgh
Supercomputing Center and the National Center for Supercomputing
Applications under grant AST070015. We would like to thank Paul Bode
for providing the N-body simulations, and Niayesh Afshordi, Keith
Vanderlinde, Paul Bode and Nicholas Hall for useful discussions.

\end{document}